\documentclass[twocolumn,prl,amsmath,amssymb,superscriptaddress]{revtex4-1}

\usepackage{graphicx}
\usepackage{dcolumn}
\usepackage{bm}
\usepackage{mathrsfs}
\usepackage{color}
\usepackage{epstopdf}
\usepackage{booktabs}
\usepackage{amsmath}
\usepackage[T1]{fontenc}
\usepackage{calligra}
\usepackage{anyfontsize}
\usepackage[normalem]{ulem}
\usepackage[colorlinks=true,linkcolor=blue,urlcolor=blue,citecolor=blue]{hyperref}

\usepackage{float}
\usepackage{physics}

\begin{document} 
\title{Fractional optical skyrmions}

\author{Yuancong Cao}
\affiliation{School of Physics, Xidian University, Xi'an, China}

\author{Ram Nandan Kumar}
\affiliation{School of Physics, University of the Witwatersrand,Johannesburg, South Africa}

\author{Hadrian Bezuidenhout}
\affiliation{School of Physics, University of the Witwatersrand,Johannesburg, South Africa}

\author{Mingjian Cheng}
\email[email:]{mjcheng@xidian.edu.cn}
\affiliation{School of Physics, Xidian University, Xi'an, China}
\email{mjcheng@xidian.edu.cn}

\author{Lixin Guo}
\affiliation{School of Physics, Xidian University, Xi'an, China}

\author{Andrew Forbes}
\email[email:]{andrew.forbes@wits.ac.za}
\affiliation{School of Physics, University of the Witwatersrand,Johannesburg, South Africa}
\email{andrew.forbes@wits.ac.za}

\date{\today}

\begin{abstract}
\noindent Optical topologies in the form of Skyrmions have attracted significant interest of late, where their integer Skyrmion number has been shown to be robust to complex media. Here we create the first fractional Skyrmions by structuring light as a vectorial superposition of non-integer orbital angular momentum.  We unravel the map structure to reveal a new phenomenon, the abrupt transition jumps in skyrmion number, which serves to reinforce the integer nature of skyrmion topologies.  Our experimental demonstration agrees well with simulation, opening a new spectrum of optical topologies to explore, with exciting possibilities in optical communication and sensing.  
\end{abstract}
 
\maketitle

\section{Introduction}

\noindent Orbital angular momentum (OAM) can be imparted to light by tailoring its spatial structure \cite{allen1992orbital}, where a phase of the form $\exp (-i \ell \phi)$ about the azimuth ($\phi$) imbues light with an OAM in the propagation direction of $\ell \hbar$ per photon for any integer $\ell$. Photonic OAM has since been a highly topical form of classical and quantum structured light \cite{forbes2021structured,forbes2025progress}, fueling many fundamental studies and applications, and has been extensively reviewed to date \cite{shen2019optical,wang2021orbital,wang2022orbital,willner2021orbital,franke202230,he2022towards,willner2021perspectives,padgett2017orbital}.  It is also possible to imbue light with a fractional OAM \cite{gotte2008light}, which has proven impactful for a range of applications \cite{zhang2022review,ni2021multidimensional}, including sensing \cite{deng2019precision,lu2025fractional,wang2025optical,cui2025recognition,berger2018spectroscopy} and communication \cite{zhang2021polarization,zhu2024security,xu2025multiplexed,zhang2025optical}.  Fractional OAM can be understood as a superposition of integer OAM, a perspective facilitated by the orthogonal nature of OAM itself.

A recent trend that is gathering in pace is to combine OAM in vectorial superpositions with polarisation to create optical Skyrmions as spin-textured fields \cite{shen2024optical}.  Such optical topologies have been created with integer topological invariants, the Skyrmion number, as single photons \cite{ma2025nanophotonic,koni2025dual} and entangled states \cite{ornelas2024non}, evanescent waves \cite{tsesses2018evanescent}, paraxial \cite{gutierrez2021optical} and non-paraxial \cite{du2019deep} light.  Interest in such topologies is fueled by their noise resilience \cite{wang2025topological,de2025quantum,ornelas2025topological,peters2025seeing,guo2025topological,Zhang2025Topological} and potential applications, ranging from communication \cite{bezuidenhout2025deep} and information processing \cite{wang2025perturbation} to metrology \cite{dai2020plasmonic}. To date, only integer topologies have been created by the superposition of integer OAM, with any deviation seen as an undesired experimental or numerical artifact.

Here we blend non-integer OAM to create the first fractional optical skyrmions, where the topological number can be tuned continuously.  We show that while fractional OAM can be understood as a superposition of integer OAM by virtue of the orthogonality of the basis, the same is not true for fractional skyrmions.  We outline how to create and understand these topologies, with experimental results in good agreement with theory.  Our work opens a new class of topologies with exciting potential as reflected by its fractional OAM counterpart, from communication to sensing. 

\begin{figure*}
    \centering
    \includegraphics[width=1\linewidth]{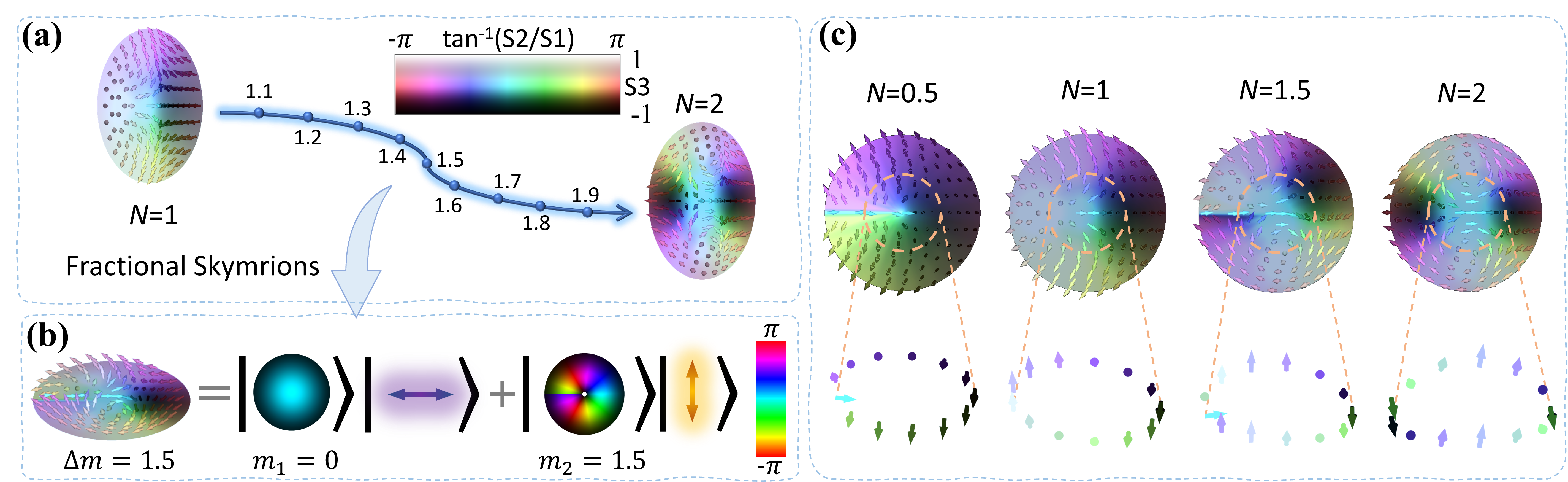}
    \caption{\textbf{Evolution and generation of fractional skyrmionic textures.} (a) Schematic illustration of fractional skyrmions acting as a bridge between integer states (exemplified by $N=1$ and $N=2$). This demonstrates a continuous evolutionary process, the mechanism of which remains largely underexplored. (b) Realization of skyrmions with arbitrary topology via the superposition of two fractional Laguerre-Gaussian (LG) beams with orthogonal polarization states (e.g., $|H\rangle$ and $|V\rangle$). The case shown corresponds to $m_1=0, m_2=1.5, \text{and } \Delta m=1.5$. (c) Comparison of polarization-texture mappings corresponding to integer and fractional order skyrmions. The integer topology (e.g., $N = 1$ and $2$) exhibits smooth, adiabatic evolution, resulting in complete coverage. However, the fractional topology (e.g., $N = 0.5$ and $1.5$) exhibits abrupt azimuthal polarization transitions, leading to incomplete polarization-texture mapping.}
    \label{fig:Concept}
\end{figure*}

\section{Concept and theory} \label{theory_sec}

Optical skyrmions are quasiparticle-like topological states characterized by spatially structured vectorial polarization textures arising from the non-separable coupling between spatial and polarization degrees of freedom. Such textures are typically realized by superposing vortex modes in linear ($\ket{H}$, $\ket{V}$) or circular ($\ket{R}$, $\ket{L}$) polarization bases. To date, optical skyrmions have predominantly been generated using integer OAM modes, leading to integer-valued topological numbers, which are invariants and illustrate smooth, adiabatic polarization evolution that fully covers the Poincaré sphere. This restriction inherently confines optical topology to discrete orders. However, fractional optical skyrmions provide a natural extension beyond this limitation. They can be generated using fractional OAM modes, which are intrinsically non-orthogonal. While fractional OAM states admit a modal decomposition into orthogonal integer OAM modes, an analogous decomposition does not extend to fractional skyrmions, since integer skyrmionic states do not form an orthogonal basis. Consequently, the topological evolution between adjacent integer skyrmionic orders is fundamentally distinct from that of fractional OAM alone. 

Studying fractional topological states is therefore essential, as they provide a continuous bridge between discrete integer topologies. In Fig.~\ref{fig:Concept}(a), we schematically illustrate the evolution of fractional topology between $N=1$ and $N=2$ with a step size of 0.1, revealing intermediate evolutionary pathways of optical topology. As shown in Fig.~\ref{fig:Concept}(b), fractional skyrmion states are formed by combining fractional OAM modes with orthogonal polarization $\ket{H}$ and $\ket{V}$ bases. In contrast, Fig.~\ref{fig:Concept}(c) highlights the distinct polarization-texture evolution for fractional and integer topologies, where integer skyrmions exhibit smooth adiabatic evolution, while fractional skyrmion states display abrupt azimuthal polarization transitions. These abrupt transitions originate from intrinsic phase discontinuities associated with fractional OAM. As a result, the Stokes vector covers only a portion of the Poincaré sphere, leading to incomplete topological wrapping. These characteristics underscore the fundamental differences between integer and fractional optical skyrmions and motivate their systematic theoretical and experimental investigation. 

Heneceforth, we define a fractional optical skyrmion beam as a coherent superposition of non-orthogonal fractional OAM modes encoded in orthogonal polarization bases, such as horizontal $\ket{H}$ and vertical $\ket{V}$ polarizations. The corresponding optical field can be written as
\begin{equation}
\ket{\psi(r,\phi)} = \alpha\, \ket{m_1}\ket{H} + \beta\, \ket{m_2}\ket{V},
\label{Eq_1}
\end{equation}
where $\ket{m_i}$ denotes a state with OAM of $m_i \hbar$ per photon.  We have without loss of generality set the first mode $\ket{m_1} \equiv \ket{0} = \sqrt{\frac{2}{\pi}}\,\frac{1}{w_0} \exp\!\left(-\frac{r^2}{w_0^2}\right)$, corresponding to a Gaussian mode carrying zero OAM.  This is valid and general as a common OAM can be factored out from both states, leaving the first term 0 and the second term a new OAM states. The fractional OAM mode (second state) is then given by $\ket{m_2} \equiv \ket{m} = \sqrt{\frac{2}{\pi\,\Gamma(|m|+1)}}\,\frac{1}{w_0} \left(\frac{\sqrt{2}\,r}{w_0}\right)^{|m|}
\exp\!\left(-\frac{r^2}{w_0^2}\right) \exp(i m \phi),$ which carries a fractional topological charge $m$. Both fields are evaluated at the beam waist ($z=0$) with radial index $p=0$. The complex coefficients $\alpha$ and $\beta$ determine the relative amplitude and phase between the two orthogonal polarization components.

The fractional OAM (non-orthogonal) modes $\ket{m}$  can be expressed through a modal decomposition in terms of known orthogonal integer-OAM basis modes as

\begin{equation}
\begin{aligned}
\ket{m} 
=\sqrt{\frac{2}{\pi\,\Gamma(|m|+1)}}\frac{1}{w_0}
\left(\frac{\sqrt{2}\, r}{w_0}\right)^{|m|}
\exp\!\left(-\frac{r^2}{w_0^2}\right) \\ \times
\sum_{n=-M}^{M} c_n(m)\,e^{i n \phi},
\label{Eq_2}
\end{aligned}
\end{equation}
where the Fourier coefficients are given by $c_n(m)=\frac{\sin[\pi(m-n)]}{\pi(m-n)}\,\exp\!\big[i\pi(m-n)\big]$. Now, using Eqs.~(\ref{Eq_1}) and~(\ref{Eq_2}), the fractional skyrmion topology state given in Eq.~(\ref{Eq_1}) can be written in a locally normalized form as

\begin{equation}
|\psi(r,\phi)\rangle
=\frac{|H\rangle+\mu(r,\phi)\,|V\rangle}
{\sqrt{1+|\mu(r,\phi)|^2}},
\label{Eq_3}
\end{equation}
where the complex field $\mu(r,\phi)=\frac{\beta}{\alpha}\,
\frac{\ket{m}}{\ket{0}}$ is given explicitly by

\begin{equation}
\mu(r,\phi)
=\frac{\beta}{\alpha}\,
\frac{1}{\sqrt{\Gamma(|m|+1)}}
\left(\frac{\sqrt{2}\, r}{w_0}\right)^{|m|}
\sum_{n=-M}^{M} c_n(m)\,e^{i n \phi}.
\label{Eq_4}
\end{equation}

The local polarization state of the fractional skyrmion beam can be represented as a point on the Poincaré sphere. This representation is obtained by evaluating the expectation value of the Pauli vector operator 
$\boldsymbol{\sigma} = (\sigma_1,\sigma_2,\sigma_3)$ 
with respect to the local normalized state $|\psi(r,\phi)\rangle$. The corresponding Stokes vector is defined as
\begin{equation}
\mathbf{S}(r,\phi)
=
\langle \psi(r,\phi) | \boldsymbol{\sigma} | \psi(r,\phi) \rangle .
\label{Eq_5}
\end{equation}

The normalized Stokes vector can be written explicitly in matrix form as
\begin{equation}
\mathbf{S}(r,\phi)
=
\frac{1}{1+|\mu(r,\phi)|^2}
\begin{pmatrix}
2\,\mathrm{Re}\!\left[\mu(r,\phi)\right] \\
2\,\mathrm{Im}\!\left[\mu(r,\phi)\right] \\
1 - |\mu(r,\phi)|^2
\end{pmatrix}.
\label{Eq_6}
\end{equation}

These normalized Stokes vector $\mathbf{S} = (S_1, S_2, S_3)$ maps an arbitrary polarization state onto the unit Poincaré sphere and completely describes the spatially varying polarization texture of the fractional optical skyrmion beam, with its topology determined by the complex field $\mu(r,\phi)$.


Through an inverse stereographic projection, the topological structure defined on the transverse plane is continuously mapped onto the unit sphere ($\mathbb{R}^2 \rightarrow \mathbb{S}^2$). The locally normalised Stokes vector then provides a simultaneous mapping from this compactified real-space sphere onto the Poincar\'e sphere, thereby expressing the spatially varying topological texture in terms of polarisation (spin) states. This composition establishes a skyrmionic texture whose topology is characterised by the associated skyrmion (or wrapping) number $N$. This quantity measures the number of times the spin texture wraps around $\mathbb{S}^2$ as the entire plane $\mathbb{R}^2$ is traversed under the mapping $\mathbb{R}^2 \rightarrow \mathbb{S}^2$, and is defined as


\begin{equation}
\scalebox{0.99}{$
N(m)
= \frac{1}{4\pi}
\int_A
\mathbf{S}\!\left[\mu(r,\phi)\right]
\cdot
\left(
\frac{\partial \mathbf{S}\!\left[\mu(r,\phi)\right]}{\partial r}
\times
\frac{\partial \mathbf{S}\!\left[\mu(r,\phi)\right]}{\partial \phi}
\right)
\, dr \, d\phi .$}
\label{Eq_7}
\end{equation}

The skyrmion number $N(m)$ provides an intuitive measure of the global spatial structure of the polarisation texture. However, an analytical evaluation of $N(m)$ is not feasible in the present framework because the fractional OAM state is expressed as a superposition of multiple orthogonal integer OAM modes, each contributing a distinct azimuthal phase. The resulting multivalued phase structure prevents a closed-form analytical integration, and therefore $N(m)$ is evaluated numerically. Unlike integer skyrmion numbers, which are directly related to the difference in OAM between the first and second terms of Eq.~\ref{Eq_1}, fractional skyrmion numbers $N(m)$ do not follow a simple linear dependence on the OAM difference. Instead, they exhibit a nonlinear relationship between the numerically evaluated fractional $N(m)$ and the corresponding OAM difference. This behavior arises from the non-orthogonal nature of fractional OAM modes and is discussed in detail in a later section.
\begin{figure}[t!]
\centering
\includegraphics[width=1\linewidth]{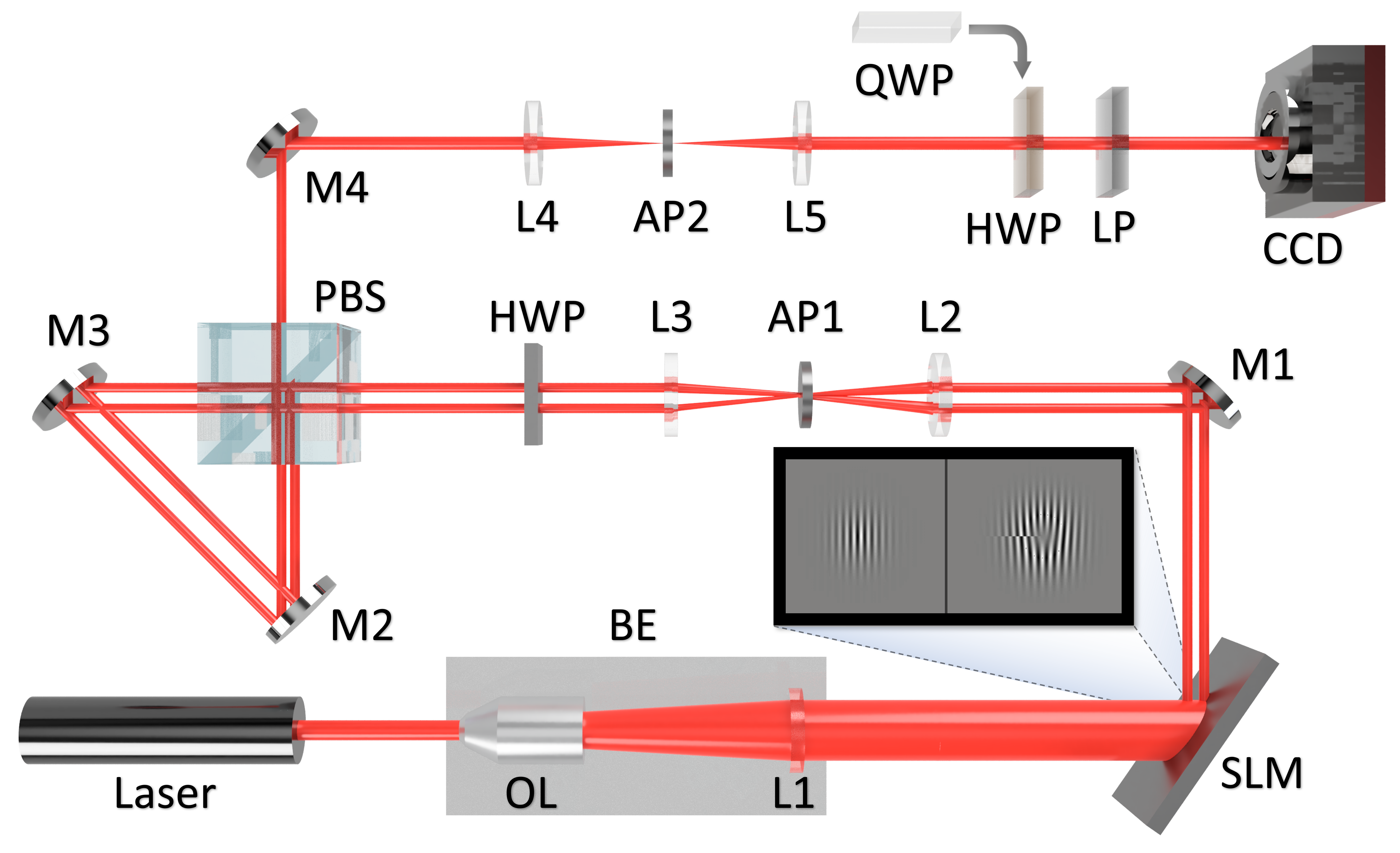}
\caption{\textbf{Experimental setup for generating and detecting fractional skyrmions.} A collimated Gaussian beam ($\lambda = 633$ nm) illuminates a spatial light modulator (SLM) encoded with a multiplexed hologram to generate two spatially separated integer or fractional LG modes. A $4f$ imaging system, formed by lenses L2 and L3 with an aperture (AP1) at the Fourier plane, filters the first diffraction orders. A half-wave plate (HWP) sets the polarization to $45^\circ$ before entering a Sagnac interferometer composed of a polarizing beam splitter (PBS) and mirrors (M2, M3). The PBS splits and subsequently recombines the orthogonal polarization components ($|H\rangle$ and $|V\rangle$), synthesizing the vector field $\ket{\psi(r,\phi)} = \ket{m_1}\ket{H} + \ket{m_2}\ket{V}$. The field is analyzed via Stokes polarimetry using a rotating quarter-wave plate (QWP) or HWP and a linear polarizer (LP), and recorded by a CCD camera. BE: beam expander; OL: objective lens; M: mirror; L: lens; CCD: charge coupled device camrea.} 
\label{fig:experimental_setup}
\end{figure} 

\section{EXPERIMENTAL CONFIGURATION AND CHARACTERIZATION}
The schematic of the experimental setup is shown in Fig.~\ref{fig:experimental_setup}. A common-path interferometric architecture was employed to generate fractional topological states with high fidelity. A Helium--Neon (He--Ne) Gaussian laser source ($\lambda = 633~\mathrm{nm}$) was converted into an approximately plane-wave input beam using a beam expander (BE) composed of a $10\times$ objective lens (numerical aperture 0.25) and a plano-convex lens (L1) with focal length $f_1 = 300~\mathrm{mm}$. This configuration ensured an approximately plane-wave illumination with uniform phase and intensity distributions across the beam profile. Subsequently, the input plane wave was incident on a spatial light modulator (SLM), where spatial multiplexing based on complex-amplitude modulation was used to generate the scalar-vortex beams. Two independent holograms carrying Laguerre--Gaussian (LG) modes with integer/fractional OAM indices $m_1$ and $m_2$ were encoded simultaneously. The emerging beams from the SLM were spatially filtered using a $4\!-\!f$ imaging system configured by lenses L2 and L3, each with focal lengths $f_2=f_3=300~\mathrm{mm}$, where an aperture (AP1) placed in the Fourier plane isolated the first diffraction orders, as shown in Fig.~\ref{fig:experimental_setup}. To facilitate vector synthesis, the resulting spatially separated, horizontally polarized beams propagated in parallel and were subsequently passed through a half-wave plate (HWP) oriented at $45^\circ$, which rotated their polarization from horizontal to a diagonal state of the form $\ket{\psi}_{1~\text{or}~2}=\frac{\ket{m_1~\text{or}~m_2}}{\sqrt{2}}(\ket{H}+\ket{V})$.

The core synthesis was performed using a modified Sagnac interferometer, formed by a polarizing beam splitter (PBS) and high-reflectivity mirrors M2 and M3. The PBS decomposed the diagonally polarized input into orthogonal horizontal ($\ket{H}$) and vertical ($\ket{V}$) components. By precisely aligning the path lengths to ensure temporal coherence and spatial overlap, we synthesized the vector vortex state via the coherent superposition $\ket\psi=\ket{m_1}\ket{H}+\ket{m_2}\ket{V}$. We again employed a $4\!-\!f$ imaging system comprising lenses L4 and L5, each with focal length $f_4 = f_5 = 300~\mathrm{mm}$, to image the synthesized fractional topological beam onto the CCD, thereby avoiding free-space propagation that could introduce aberrations.

\begin{figure*}[t!]
    \centering
    \includegraphics[width=\linewidth]{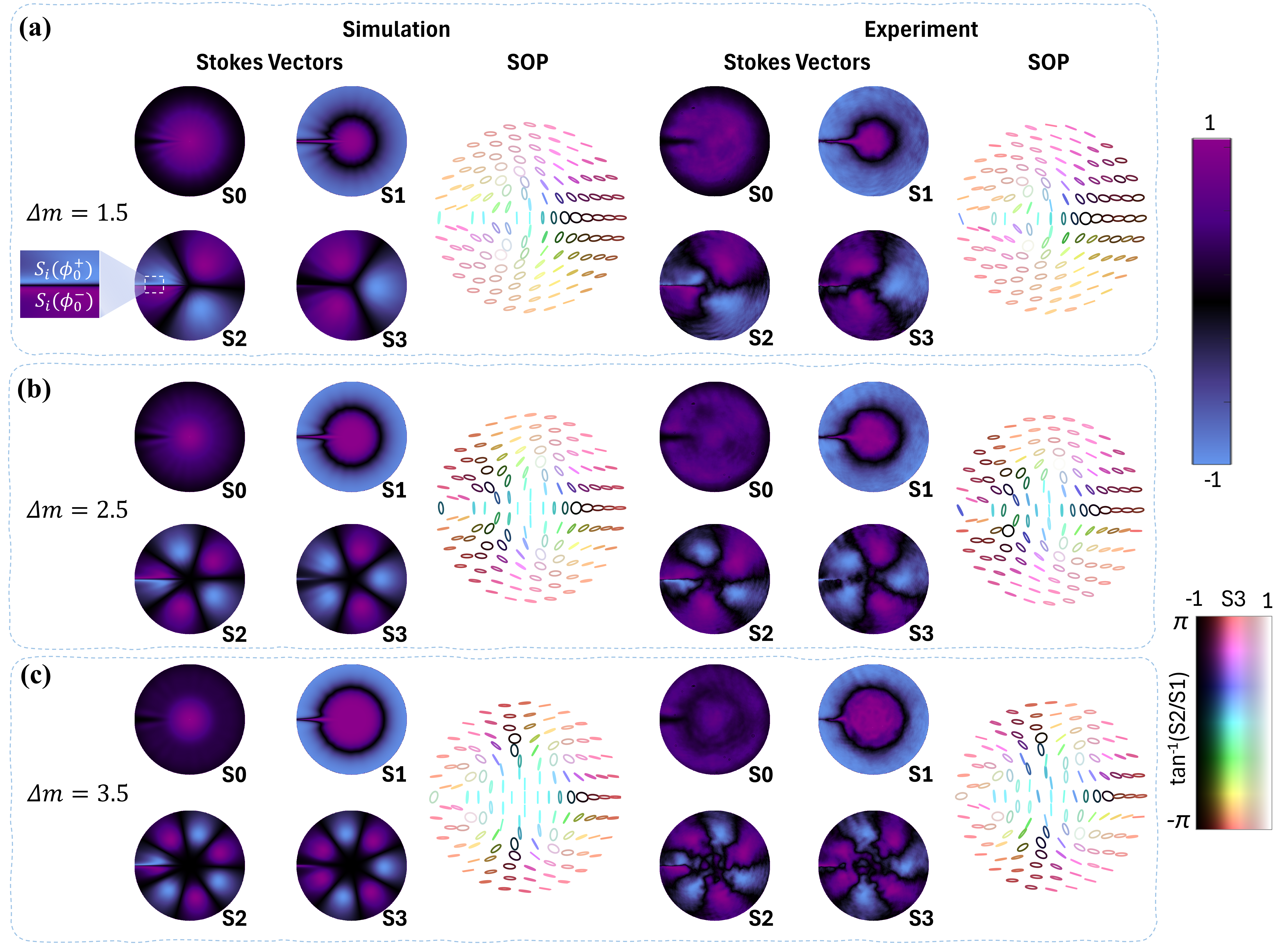}
    \caption{\textbf{Stokes fields and state of polarization of fractional topologies.} 
    The Stokes parameters ($S_0$, $S_1$, $S_2$, and $S_3$) and the corresponding states of polarization (SOP) for fractional topological states described by Eq.~\ref{Eq_1} are shown for (a) $m_1 = 0$ and $m_2 = 1.5$ ($\Delta m = 1.5$), (b) $m_1 = 0$ and $m_2 = 2.5$ ($\Delta m = 2.5$), and (c) $m_1 = 0$ and $m_2 = 3.5$ ($\Delta m = 3.5$). The left and right columns present the numerically simulated and experimentally measured results, respectively. By precisely controlling the relative phases ($\alpha$ and $\beta$) in Eq.~\ref{Eq_1}, excellent agreement between simulations and experiments is achieved. Owing to the fractional topological charge $m_2$, a discontinuity appears at the azimuthal angle $\phi = \phi_0$ in the Stokes fields, such that $S_i(\phi_0^-) \neq S_i(\phi_0^+)$, which leads to a sudden rotation of the polarization ellipse in the SOP distributions.}    
    \label{Fig_3}
\end{figure*}

To characterize the topological structure of the synthesized field, full Stokes polarimetry was employed. A polarization state analyzer, consisting of a rotating quarter-wave plate (QWP) or half-wave plate (HWP) and a fixed linear polarizer (LP), was placed before the CCD detector, as shown in Fig.~\ref{fig:experimental_setup}. A combination of an LP and an HWP was used to acquire the intensity projections corresponding to linear polarization components, while an LP in combination with a QWP was employed to measure the circular polarization projections. Thus, the Stokes parameters were experimentally determined by measuring six polarization-resolved intensity projections, given as

\begin{equation}
\mathbf{S} =
\begin{pmatrix}
S_0 \\[2pt]
S_1 \\[2pt]
S_2 \\[2pt]
S_3
\end{pmatrix}
=\begin{pmatrix}
I_H + I_V \\[4pt]
I_H - I_V \\[4pt]
I_D - I_A \\[4pt]
I_R - I_L
\end{pmatrix}_{exp}
=
\begin{pmatrix}
|E_x|^2 + |E_y|^2 \\[4pt]
|E_x|^2 - |E_y|^2 \\[4pt]
2\,\mathrm{Re}(E_x E_y^{*}) \\[4pt]
2\,\mathrm{Im}(E_x E_y^{*})
\end{pmatrix}_{theory}.
\label{eq:stokes}
\end{equation}

Here, the subscripts $H$, $V$, $D$, $A$, $R$, and $L$ represent horizontal, vertical, diagonal, antidiagonal, right-circular, and left-circular polarizations, respectively. According to Eqs.~\ref{Eq_6} and \ref{Eq_7}, the locally normalized Stokes parameters $S_i$ were required and were computed from the experimentally obtained Stokes parameters $S_i$ [Eq.~\ref{eq:stokes}] as $S_i = \frac{S_i}{\sqrt{S_1^2 + S_2^2 + S_3^2}}$. However, in theory, the locally normalized Stokes parameters $S_i$ were calculated using Eq.~\ref{Eq_6}, corresponding to the experimentally generated fractional topological states. Finally, these locally normalized Stokes parameters were used to numerically compute the skyrmion (wrapping) numbers given in Eq.~\ref{Eq_7}, both theoretically and experimentally. The theoretical results reveal the characteristic intensity discontinuities and polarization singularities associated with fractional skyrmionic textures. In the following section, a comparison between the theoretical and experimental results is presented.

\begin{figure*}
    \centering
    \includegraphics[width=1\linewidth]{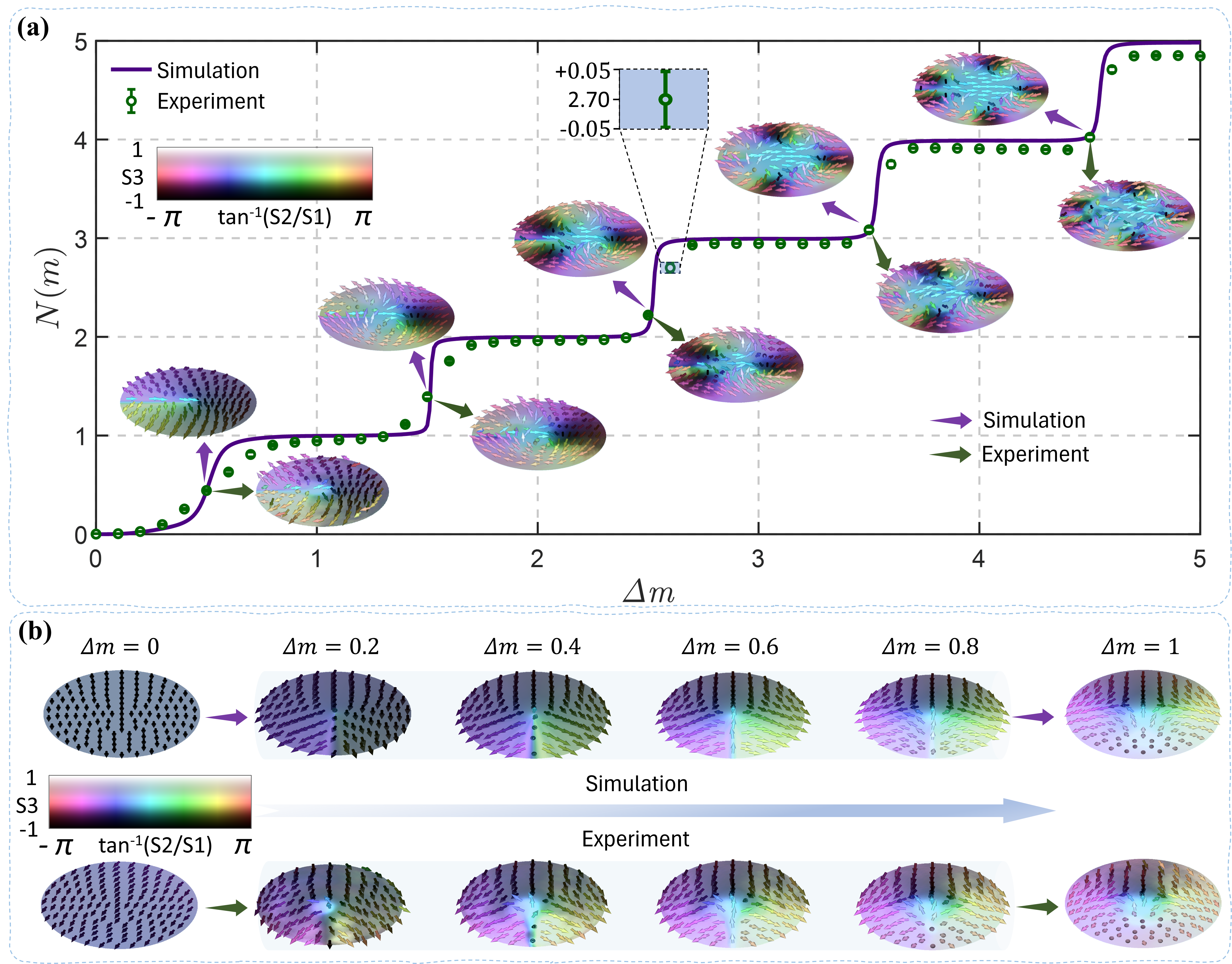}
    \caption{\textbf{Fractional topological Skyrmion numbers and evolution of topological textures.} (a) Evolution of Skyrmion numbers within the interval $\Delta m \in [0, 5]$. The purple solid lines represent simulation results, while the green data points (with a step size of 0.1) denote the mean values of 15 experimental measurements, with error bars indicating the standard deviation. The topological textures at half-integer OAM differences ($\Delta m = 0.5, 1.5, 2.5, 3.5,$ and $4.5$) are shown, where the purple and green arrows represent the simulation and experimental results, respectively. The fractional skyrmion number exhibits a nonlinear variation between adjacent integer topological charges. Specifically, this trend is characterized by rapid changes in the vicinity of half-integer values, while remaining relatively gradual near integer regions. Furthermore, this evolutionary behavior becomes increasingly pronounced as $\Delta m$ increases. The experimental results are in good agreement with the simulation curves, verifying the nonlinear evolutionary characteristics of fractional Skyrmions. (b)The continuous evolution of the topological texture from$\Delta m$ = 0 to $\Delta m$= 1 (step by 0.2). The top row (simulation) and bottom row (experiment) visualize the bifurcation of the Skyrmion core
mediated by the azimuthal discontinuity. }
    \label{Fig_4}
\end{figure*}

\section{Results and Discussion}
\begin{figure}[t!]
\centering
\includegraphics[width=\linewidth]{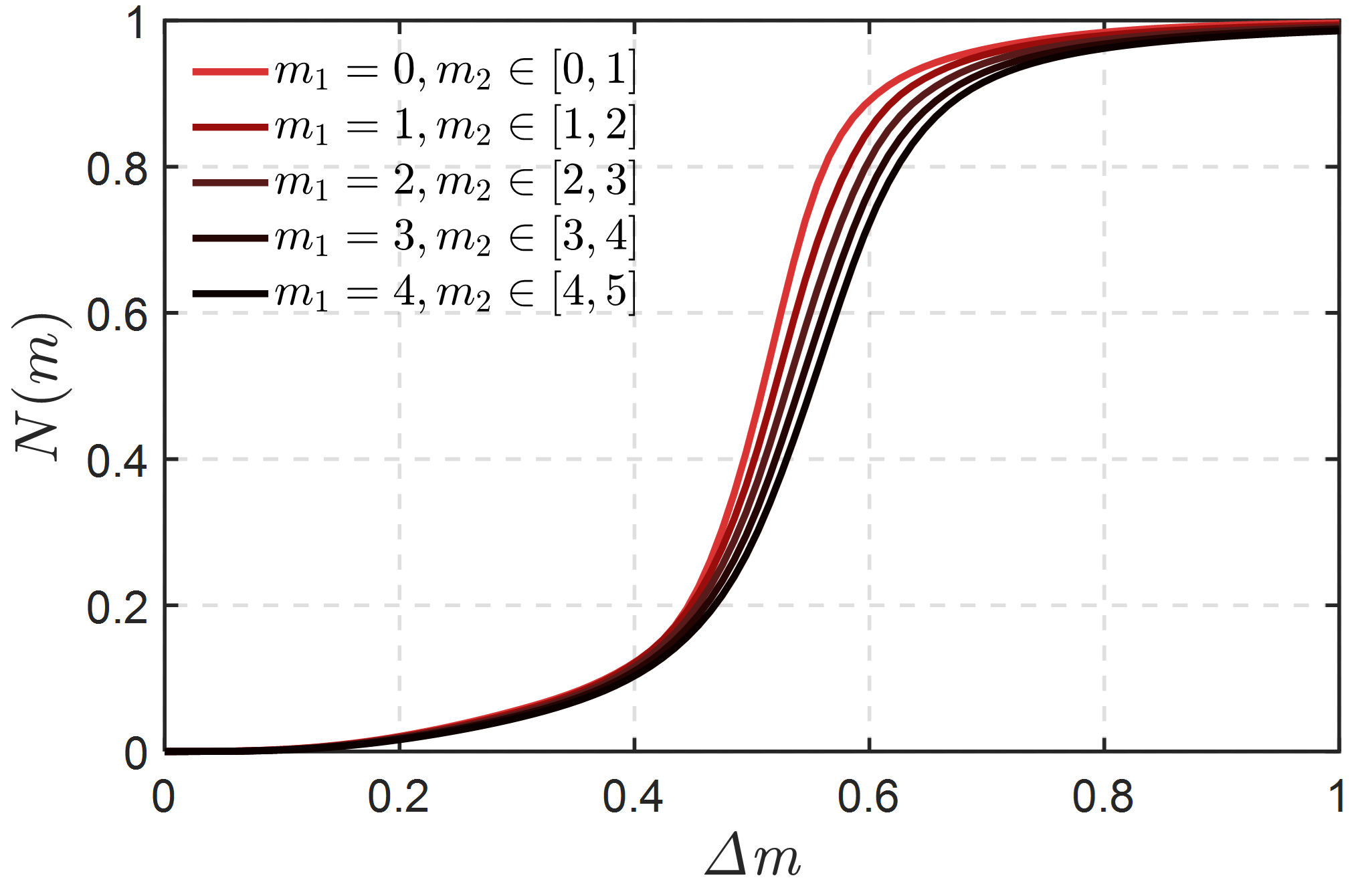}
\caption{\textbf{Evolution of fractional skyrmions under different OAM configurations.} Evolutionary trajectories are shown for five OAM combinations within the interval $\Delta m \in [0, 1]$: $(m_1=0, m_2 \in [0,1])$, $(m_1=1, m_2 \in [1,2])$, $(m_1=2, m_2 \in [2,3])$, $(m_1=3, m_2 \in [3,4])$, and $(m_1=4, m_2 \in [4,5])$. Higher OAM configurations lead to a systematic shift of the transition curves toward larger $\Delta m$. The non-linear profile of these curves reveals the fundamental nonlinear dynamics of fractional topological evolution.}
\label{Fig_5}
\end{figure}
Integer and non-integer skyrmionic topological beams can simultaneously encode multiple degrees of freedom (DoFs) of light. In this work, we focused on the non-separable coupling between polarization and an azimuthally varying phase structure, as described in Eq.~\ref{Eq_1}. A non-zero skyrmion number arises from the coherent superposition of scalar-vortex modes in orthogonal polarization bases—either linear ($\ket{H}$, $\ket{V}$) or circular ($\ket{R}$, $\ket{L}$)—with a finite azimuthal phase difference between the constituent spatial modes. The skyrmion number $N$, also referred to as the wrapping number, characterizes the number of times the polarization texture wraps the Poincar\'e sphere. For integer skyrmionic states, the wrapping number $N$ signifies that the polarization texture completely covers the Poincar\'e sphere an integer number of times and is linearly related to the OAM mode difference $\Delta m = m_2 - m_1$ between the constituent modes. In contrast, the fractional skyrmion number $N(m)$, given in Eq.~\ref{Eq_7}, does not exhibit a linear dependence on the OAM mode difference and corresponds to a polarization texture that wraps only a portion of the Poincar\'e sphere rather than covering it entirely, as will be discussed later in this section. The fractional skyrmion number $N(m)$ was evaluated using a mode-decomposition approach, with the detailed procedure described in the theory Sec.~\ref {theory_sec}. Notably, fractional skyrmionic topological configurations enable a continuous evolution of the topological landscape between discrete integer topological orders. For simplicity, the first spatial mode ($\ket{m_1}$) was chosen as the fundamental Gaussian mode ($m_1=0$), while the second spatial mode ($\ket{m_2}$) consisted of the fractional OAM mode associated with the orthogonal polarization bases $\ket{H}$ and $\ket{V}$, respectively, as defined in Eq.~\ref{Eq_1}. 
We first describe the Stokes vectors and the corresponding states of polarization (SOP) for half-integer modes, followed by a systematic analysis of fractional topological states and the evolution of their polarization textures.

\subsection{Stokes polarimetry of fractional skyrmion fields}

To quantitatively characterize the SOP, or spin texture, we employed the experimental setup shown in Fig.~\ref{fig:experimental_setup} to perform polarization-resolved intensity projection measurements, namely $I_H$, $I_V$, $I_D$, $I_A$, $I_R$, and $I_L$. From these measured intensities, the Stokes parameters $S_0$, $S_1$, $S_2$, and $S_3$ were reconstructed using Eq.~\ref{eq:stokes}. Figure~\ref{Fig_3} shows the numerically simulated and experimentally measured Stokes parameters, along with the corresponding SOP distributions, for a half-integer topological state for the cases of (a) $\Delta m = 1.5$, (b) $\Delta m = 2.5$ and (c) $\Delta m = 3.5$, all under the condition that $m_1 = 0$. The analytical expressions of $S_1$ and $S_2$ explicitly contain the real and imaginary parts of the complex field $\mu(r,\phi)$ (see Eq.~\ref{Eq_6}). For fractional OAM modes, $\mu(r,\phi)$ is multivalued and satisfies $\mu(r,\phi + 2\pi) = \mu(r,\phi)e^{i2\pi m} \neq \mu(r,\phi)$, which necessitates the presence of a branch cut at a specific azimuthal angle $\phi = \phi_0$. Consequently, the left- and right-hand limits of $\mu(r,\phi)$ across the branch cut are unequal, i.e., $\lim_{\phi \to \phi_0^-} \mu(r,\phi) \neq \lim_{\phi \to \phi_0^+} \mu(r,\phi)$. This discontinuity is directly transferred to the linear Stokes parameters, leading to $S_i(\phi_0^-) \neq S_i(\phi_0^+)$, as clearly shown in Fig.~\ref{Fig_3}(a). As a result, the azimuthal symmetry of the Stokes fields is broken, giving rise to a sudden rotation of the polarization ellipse in the SOP plots. In contrast, the $S_3$ parameter represents the degree of circular polarization, corresponding to right-handed ($+1$) and left-handed ($-1$) helicities, depicted by deep purple and cornflower blue colormaps, respectively, in the Stokes vector plots.  This behavior is further corroborated in the SOP distributions, where white and black polarization ellipses indicate positive and negative helicities, respectively.

Similarly, Figs~\ref{Fig_3}(b) and~\ref{Fig_3}(c) illustrate the Stokes polarimetry results for higher-order fractional topological states. In these cases, increasingly complex polarization textures emerge due to the higher-order half-integer fractional OAM modes. Overall, the Stokes polarimetry results show excellent qualitative agreement between numerical simulations and experimental measurements of the Stokes parameters and the associated SOP distributions of fractional topological states.

\subsection{Abrupt evolution of fractional topology}

In Fig.~\ref{Fig_4}(a), we present the numerically calculated and experimentally measured fractional skyrmion number $N(m)$ as a function of the OAM mode difference $\Delta m$ between the constituent spatial modes defined in Eq.~\ref{Eq_1}. The evolution of the fractional skyrmion number is shown over the interval $\Delta m \in [0,5]$ with a step size of $\Delta m=0.1$. To elucidate the topological transitions between successive integer values, we observe both smooth and abrupt transitions in fractional topology. For lower values of $\Delta m$, the skyrmion number $N(m)$ varies smoothly, continuously, and nonlinearly, indicating a gradual evolution of fractional topology in this regime. This behavior can be attributed to the intrinsically nonlinear nature of the complex field $\mu(r,\phi)$ for fractional OAM values. The corresponding topological transitions occur through inflection points in the $N(m)$ curve, which are located close to the half-integer values of $\Delta m$. In contrast, for higher values of $\Delta m$, the skyrmion number $N(m)$ evolves abruptly, indicating a rapid transition of fractional topological states toward integer values. This behavior can be understood by examining the weighting coefficients arising from the modal decomposition of fractional OAM modes. According to the modal superposition principle, when the fractional OAM value lies within the interval $m_2 \in [\mathbb{Z}-0.5,\,\mathbb{Z}+0.5]$, the corresponding weighting coefficient exhibits a pronounced peak at the nearest integer $\mathbb{Z}$, while rapidly diminishing for adjacent integer orders. Consequently, for $\Delta m \in [\mathbb{Z}-0.5,\,\mathbb{Z}+0.5]$, the fractional skyrmion number $N(m)$ remains pinned close to the integer value $\mathbb{Z}$, reflecting the dominant contribution of the integer OAM component in this range. 

Note that at the half-integer values of $\Delta m$, namely $0.5$, $1.5$, $2.5$, $3.5$, and $4.5$ the experimentally measured and numerically simulated polarization textures are also shown in Fig.~\ref{Fig_4}(a). In these plots, vectors pointing upward and downward represent helicities of $+1$ and $-1$, respectively, while the azimuthal color variation encodes the local orientation angle of the polarization ellipse. Together, these features visualize the fractional wrapping of the polarization texture on the Poincar\'e sphere. Owing to the discontinuities in the Stokes parameters discussed earlier, abrupt changes in the polarization orientation are observed at the branch-cut locations $\phi=\phi_0$, as highlighted in Figs.~\ref{Fig_4}(a)--(b). Figure~\ref{Fig_4}(b) further illustrates the continuous evolution of the polarization texture as $\Delta m$ is varied from 0 to 1 in steps of 0.2. Both experimental and numerical results reveal that the degree of coverage of the Poincar\'e sphere evolves continuously but nonlinearly with $\Delta m$. In this sense, the fractional skyrmion number $N(m)$ acts as a bridge connecting discrete integer topological orders. This continuous evolution establishes a robust link between adjacent integer topologies and reveals the intrinsic dynamical behavior of fractional optical skyrmions.

\subsection{Mode-dependent topological transitions}
In the preceding analysis, we demonstrated that the fractional skyrmion number $N(m)$ evolves abruptly between adjacent integer topologies. However, since fractional topological states can be generated from multiple OAM base combinations, it is essential to determine whether this evolution is universal or depends on the specific OAM configuration $[m_1, m_2]$ that yields the same topological difference $\Delta m$. To address this question, we extend our investigation to examine the mode-dependent nonlinear evolution of fractional topology within the fixed interval $\Delta m \in [0,1]$ for different OAM base combinations $[m_1, m_2]$. The resulting evolutionary patterns of fractional skyrmions are presented in Fig.~\ref{Fig_5}. In particular, as the OAM configuration is shifted to higher values, such as $(m_1=1,\, m_2 \in [1,2])$, $(m_1=2,\, m_2 \in [2,3])$, $(m_1=3,\, m_2 \in [3,4])$, and $(m_1=4,\, m_2 \in [4,5])$, the corresponding topological transition curves gradually shift toward larger values of $\Delta m$. This shift leads to an increase in the inflection point of $\Delta m$ and, consequently, the fractional skyrmion number $N(m)$ systematically decreases for higher OAM configurations $[m_1,\, m_2]$. Notably, the most significant deviations in $N(m)$ occur during the intermediate stages of $\Delta m$. This behavior can be attributed to the gradual growing nature of the phase singularities associated with the individual modes $m_1$ and $m_2$. In particular, imperfect spatial matching between the two constituent modes reduces the effective coupling strength between the polarization and spatial DoFs, thereby lowering the resulting fractional skyrmion number. 

\section{Conclusion}

We have demonstrated a robust framework for generating and controlling fractional optical skyrmions mediated by fractional OAM modes. Because fractional OAM states are non-orthogonal, they can be expressed as linear superpositions of orthogonal integer OAM modes via modal decomposition; however, this approach does not directly extend to fractional skyrmions, as integer skyrmionic states do not form an orthogonal basis. Our experimental and theoretical results show that the fractional skyrmion number $N(m)$ evolves between adjacent integers through an abrupt change, thereby bridging discrete integer topological orders and reinforcing the integer nature. This behavior originates from the intrinsic phase discontinuities of fractional OAM modes, which are inherited by the Stokes fields. Consequently, abrupt changes in ellipticity appear in the SOP and polarization texture at the branch cut, reflecting the incomplete wrapping of the spin texture on the Poincaré sphere. Furthermore, we numerically examined the fractional topological evolution within the interval $\Delta m \in [0,1]$ for different OAM configurations $[m_1, m_2]$. During the intermediate stages of $\Delta m$, we observed a systematic reduction of $N(m)$ for higher OAM configurations. This reduction arises from the gradual growth of phase singularities associated with the individual modes $m_1$ and $m_2$, which weakens the effective coupling between the spatial and polarization DoFs and thereby lowers the resulting $N(m)$. Future studies may further exploit this evolution to engineer tailored fractional topological responses. Overall, our work establishes a generalized framework for optical topology beyond integer quantization, where the emergence of continuous topological DoFs may enable high-density optical information encoding, and sensitive topological detectors exploiting the rapid rate of change in the continuous slopes of topological transitions.

\section{AUTHOR DECLARATIONS}

\textbf{Conflict of Interest}

The authors declare no conflicts of interest.\\

\textbf{Data Availability}

The data that support the findings of this study are available from the corresponding author upon reasonable request.\\

\textbf{Author Contributions}

All authors take full responsibility for the content of this manuscript, have reviewed the results, and have approved the final manuscript for submission.\\


\begin{thebibliography}{39}%
\makeatletter
\providecommand \@ifxundefined [1]{%
 \@ifx{#1\undefined}
}%
\providecommand \@ifnum [1]{%
 \ifnum #1\expandafter \@firstoftwo
 \else \expandafter \@secondoftwo
 \fi
}%
\providecommand \@ifx [1]{%
 \ifx #1\expandafter \@firstoftwo
 \else \expandafter \@secondoftwo
 \fi
}%
\providecommand \natexlab [1]{#1}%
\providecommand \enquote  [1]{``#1''}%
\providecommand \bibnamefont  [1]{#1}%
\providecommand \bibfnamefont [1]{#1}%
\providecommand \citenamefont [1]{#1}%
\providecommand \href@noop [0]{\@secondoftwo}%
\providecommand \href [0]{\begingroup \@sanitize@url \@href}%
\providecommand \@href[1]{\@@startlink{#1}\@@href}%
\providecommand \@@href[1]{\endgroup#1\@@endlink}%
\providecommand \@sanitize@url [0]{\catcode `\\12\catcode `\$12\catcode `\&12\catcode `\#12\catcode `\^12\catcode `\_12\catcode `\%12\relax}%
\providecommand \@@startlink[1]{}%
\providecommand \@@endlink[0]{}%
\providecommand \url  [0]{\begingroup\@sanitize@url \@url }%
\providecommand \@url [1]{\endgroup\@href {#1}{\urlprefix }}%
\providecommand \urlprefix  [0]{URL }%
\providecommand \Eprint [0]{\href }%
\providecommand \doibase [0]{http://dx.doi.org/}%
\providecommand \selectlanguage [0]{\@gobble}%
\providecommand \bibinfo  [0]{\@secondoftwo}%
\providecommand \bibfield  [0]{\@secondoftwo}%
\providecommand \translation [1]{[#1]}%
\providecommand \BibitemOpen [0]{}%
\providecommand \bibitemStop [0]{}%
\providecommand \bibitemNoStop [0]{.\EOS\space}%
\providecommand \EOS [0]{\spacefactor3000\relax}%
\providecommand \BibitemShut  [1]{\csname bibitem#1\endcsname}%
\let\auto@bib@innerbib\@empty
\bibitem [{\citenamefont {Allen}\ \emph {et~al.}(1992)\citenamefont {Allen}, \citenamefont {Beijersbergen}, \citenamefont {Spreeuw},\ and\ \citenamefont {Woerdman}}]{allen1992orbital}%
  \BibitemOpen
  \bibfield  {author} {\bibinfo {author} {\bibfnamefont {L.}~\bibnamefont {Allen}}, \bibinfo {author} {\bibfnamefont {M.~W.}\ \bibnamefont {Beijersbergen}}, \bibinfo {author} {\bibfnamefont {R.}~\bibnamefont {Spreeuw}}, \ and\ \bibinfo {author} {\bibfnamefont {J.}~\bibnamefont {Woerdman}},\ }\href@noop {} {\bibfield  {journal} {\bibinfo  {journal} {Physical Review A}\ }\textbf {\bibinfo {volume} {45}},\ \bibinfo {pages} {8185} (\bibinfo {year} {1992})}\BibitemShut {NoStop}%
\bibitem [{\citenamefont {Forbes}\ \emph {et~al.}(2021)\citenamefont {Forbes}, \citenamefont {De~Oliveira},\ and\ \citenamefont {Dennis}}]{forbes2021structured}%
  \BibitemOpen
  \bibfield  {author} {\bibinfo {author} {\bibfnamefont {A.}~\bibnamefont {Forbes}}, \bibinfo {author} {\bibfnamefont {M.}~\bibnamefont {De~Oliveira}}, \ and\ \bibinfo {author} {\bibfnamefont {M.~R.}\ \bibnamefont {Dennis}},\ }\href@noop {} {\bibfield  {journal} {\bibinfo  {journal} {Nature Photonics}\ }\textbf {\bibinfo {volume} {15}},\ \bibinfo {pages} {253} (\bibinfo {year} {2021})}\BibitemShut {NoStop}%
\bibitem [{\citenamefont {Forbes}\ \emph {et~al.}(2025)\citenamefont {Forbes}, \citenamefont {Nothlawala},\ and\ \citenamefont {Vall{\'e}s}}]{forbes2025progress}%
  \BibitemOpen
  \bibfield  {author} {\bibinfo {author} {\bibfnamefont {A.}~\bibnamefont {Forbes}}, \bibinfo {author} {\bibfnamefont {F.}~\bibnamefont {Nothlawala}}, \ and\ \bibinfo {author} {\bibfnamefont {A.}~\bibnamefont {Vall{\'e}s}},\ }\href@noop {} {\bibfield  {journal} {\bibinfo  {journal} {Nature Photonics}\ }\textbf {\bibinfo {volume} {19}},\ \bibinfo {pages} {1291} (\bibinfo {year} {2025})}\BibitemShut {NoStop}%
\bibitem [{\citenamefont {Shen}\ \emph {et~al.}(2019)\citenamefont {Shen}, \citenamefont {Wang}, \citenamefont {Xie}, \citenamefont {Min}, \citenamefont {Fu}, \citenamefont {Liu}, \citenamefont {Gong},\ and\ \citenamefont {Yuan}}]{shen2019optical}%
  \BibitemOpen
  \bibfield  {author} {\bibinfo {author} {\bibfnamefont {Y.}~\bibnamefont {Shen}}, \bibinfo {author} {\bibfnamefont {X.}~\bibnamefont {Wang}}, \bibinfo {author} {\bibfnamefont {Z.}~\bibnamefont {Xie}}, \bibinfo {author} {\bibfnamefont {C.}~\bibnamefont {Min}}, \bibinfo {author} {\bibfnamefont {X.}~\bibnamefont {Fu}}, \bibinfo {author} {\bibfnamefont {Q.}~\bibnamefont {Liu}}, \bibinfo {author} {\bibfnamefont {M.}~\bibnamefont {Gong}}, \ and\ \bibinfo {author} {\bibfnamefont {X.}~\bibnamefont {Yuan}},\ }\href@noop {} {\bibfield  {journal} {\bibinfo  {journal} {Light: Science \& Applications}\ }\textbf {\bibinfo {volume} {8}},\ \bibinfo {pages} {1} (\bibinfo {year} {2019})}\BibitemShut {NoStop}%
\bibitem [{\citenamefont {Wang}\ \emph {et~al.}(2021)\citenamefont {Wang}, \citenamefont {Chen},\ and\ \citenamefont {Liu}}]{wang2021orbital}%
  \BibitemOpen
  \bibfield  {author} {\bibinfo {author} {\bibfnamefont {J.}~\bibnamefont {Wang}}, \bibinfo {author} {\bibfnamefont {S.}~\bibnamefont {Chen}}, \ and\ \bibinfo {author} {\bibfnamefont {J.}~\bibnamefont {Liu}},\ }\href@noop {} {\bibfield  {journal} {\bibinfo  {journal} {APL Photonics}\ }\textbf {\bibinfo {volume} {6}},\ \bibinfo {pages} {060804} (\bibinfo {year} {2021})}\BibitemShut {NoStop}%
\bibitem [{\citenamefont {Wang}\ \emph {et~al.}(2022)\citenamefont {Wang}, \citenamefont {Liu}, \citenamefont {Li}, \citenamefont {Zhao}, \citenamefont {Du},\ and\ \citenamefont {Zhu}}]{wang2022orbital}%
  \BibitemOpen
  \bibfield  {author} {\bibinfo {author} {\bibfnamefont {J.}~\bibnamefont {Wang}}, \bibinfo {author} {\bibfnamefont {J.}~\bibnamefont {Liu}}, \bibinfo {author} {\bibfnamefont {S.}~\bibnamefont {Li}}, \bibinfo {author} {\bibfnamefont {Y.}~\bibnamefont {Zhao}}, \bibinfo {author} {\bibfnamefont {J.}~\bibnamefont {Du}}, \ and\ \bibinfo {author} {\bibfnamefont {L.}~\bibnamefont {Zhu}},\ }\href@noop {} {\bibfield  {journal} {\bibinfo  {journal} {Nanophotonics}\ }\textbf {\bibinfo {volume} {11}},\ \bibinfo {pages} {645} (\bibinfo {year} {2022})}\BibitemShut {NoStop}%
\bibitem [{\citenamefont {Willner}\ \emph {et~al.}(2021{\natexlab{a}})\citenamefont {Willner}, \citenamefont {Pang}, \citenamefont {Song}, \citenamefont {Zou},\ and\ \citenamefont {Zhou}}]{willner2021orbital}%
  \BibitemOpen
  \bibfield  {author} {\bibinfo {author} {\bibfnamefont {A.~E.}\ \bibnamefont {Willner}}, \bibinfo {author} {\bibfnamefont {K.}~\bibnamefont {Pang}}, \bibinfo {author} {\bibfnamefont {H.}~\bibnamefont {Song}}, \bibinfo {author} {\bibfnamefont {K.}~\bibnamefont {Zou}}, \ and\ \bibinfo {author} {\bibfnamefont {H.}~\bibnamefont {Zhou}},\ }\href@noop {} {\bibfield  {journal} {\bibinfo  {journal} {Applied Physics Reviews}\ }\textbf {\bibinfo {volume} {8}},\ \bibinfo {pages} {041312} (\bibinfo {year} {2021}{\natexlab{a}})}\BibitemShut {NoStop}%
\bibitem [{\citenamefont {Franke-Arnold}(2022)}]{franke202230}%
  \BibitemOpen
  \bibfield  {author} {\bibinfo {author} {\bibfnamefont {S.}~\bibnamefont {Franke-Arnold}},\ }\href@noop {} {\bibfield  {journal} {\bibinfo  {journal} {Nature Reviews Physics}\ }\textbf {\bibinfo {volume} {4}},\ \bibinfo {pages} {361} (\bibinfo {year} {2022})}\BibitemShut {NoStop}%
\bibitem [{\citenamefont {He}\ \emph {et~al.}(2022)\citenamefont {He}, \citenamefont {Shen},\ and\ \citenamefont {Forbes}}]{he2022towards}%
  \BibitemOpen
  \bibfield  {author} {\bibinfo {author} {\bibfnamefont {C.}~\bibnamefont {He}}, \bibinfo {author} {\bibfnamefont {Y.}~\bibnamefont {Shen}}, \ and\ \bibinfo {author} {\bibfnamefont {A.}~\bibnamefont {Forbes}},\ }\href@noop {} {\bibfield  {journal} {\bibinfo  {journal} {Light: Science \& Applications}\ }\textbf {\bibinfo {volume} {11}},\ \bibinfo {pages} {1} (\bibinfo {year} {2022})}\BibitemShut {NoStop}%
\bibitem [{\citenamefont {Willner}\ \emph {et~al.}(2021{\natexlab{b}})\citenamefont {Willner}, \citenamefont {Zhao}, \citenamefont {Liu}, \citenamefont {Zhang}, \citenamefont {Song}, \citenamefont {Pang}, \citenamefont {Manukyan}, \citenamefont {Song}, \citenamefont {Su}, \citenamefont {Xie} \emph {et~al.}}]{willner2021perspectives}%
  \BibitemOpen
  \bibfield  {author} {\bibinfo {author} {\bibfnamefont {A.~E.}\ \bibnamefont {Willner}}, \bibinfo {author} {\bibfnamefont {Z.}~\bibnamefont {Zhao}}, \bibinfo {author} {\bibfnamefont {C.}~\bibnamefont {Liu}}, \bibinfo {author} {\bibfnamefont {R.}~\bibnamefont {Zhang}}, \bibinfo {author} {\bibfnamefont {H.}~\bibnamefont {Song}}, \bibinfo {author} {\bibfnamefont {K.}~\bibnamefont {Pang}}, \bibinfo {author} {\bibfnamefont {K.}~\bibnamefont {Manukyan}}, \bibinfo {author} {\bibfnamefont {H.}~\bibnamefont {Song}}, \bibinfo {author} {\bibfnamefont {X.}~\bibnamefont {Su}}, \bibinfo {author} {\bibfnamefont {G.}~\bibnamefont {Xie}},  \emph {et~al.},\ }\href@noop {} {\bibfield  {journal} {\bibinfo  {journal} {Apl Photonics}\ }\textbf {\bibinfo {volume} {6}},\ \bibinfo {pages} {030901} (\bibinfo {year} {2021}{\natexlab{b}})}\BibitemShut {NoStop}%
\bibitem [{\citenamefont {Padgett}(2017)}]{padgett2017orbital}%
  \BibitemOpen
  \bibfield  {author} {\bibinfo {author} {\bibfnamefont {M.~J.}\ \bibnamefont {Padgett}},\ }\href@noop {} {\bibfield  {journal} {\bibinfo  {journal} {Optics Express}\ }\textbf {\bibinfo {volume} {25}},\ \bibinfo {pages} {11265} (\bibinfo {year} {2017})}\BibitemShut {NoStop}%
\bibitem [{\citenamefont {G{\"o}tte}\ \emph {et~al.}(2008)\citenamefont {G{\"o}tte}, \citenamefont {O’Holleran}, \citenamefont {Preece}, \citenamefont {Flossmann}, \citenamefont {Franke-Arnold}, \citenamefont {Barnett},\ and\ \citenamefont {Padgett}}]{gotte2008light}%
  \BibitemOpen
  \bibfield  {author} {\bibinfo {author} {\bibfnamefont {J.~B.}\ \bibnamefont {G{\"o}tte}}, \bibinfo {author} {\bibfnamefont {K.}~\bibnamefont {O’Holleran}}, \bibinfo {author} {\bibfnamefont {D.}~\bibnamefont {Preece}}, \bibinfo {author} {\bibfnamefont {F.}~\bibnamefont {Flossmann}}, \bibinfo {author} {\bibfnamefont {S.}~\bibnamefont {Franke-Arnold}}, \bibinfo {author} {\bibfnamefont {S.~M.}\ \bibnamefont {Barnett}}, \ and\ \bibinfo {author} {\bibfnamefont {M.~J.}\ \bibnamefont {Padgett}},\ }\href@noop {} {\bibfield  {journal} {\bibinfo  {journal} {Optics express}\ }\textbf {\bibinfo {volume} {16}},\ \bibinfo {pages} {993} (\bibinfo {year} {2008})}\BibitemShut {NoStop}%
\bibitem [{\citenamefont {Zhang}\ \emph {et~al.}(2022)\citenamefont {Zhang}, \citenamefont {Zeng}, \citenamefont {Lu}, \citenamefont {Wang}, \citenamefont {Zhao},\ and\ \citenamefont {Cai}}]{zhang2022review}%
  \BibitemOpen
  \bibfield  {author} {\bibinfo {author} {\bibfnamefont {H.}~\bibnamefont {Zhang}}, \bibinfo {author} {\bibfnamefont {J.}~\bibnamefont {Zeng}}, \bibinfo {author} {\bibfnamefont {X.}~\bibnamefont {Lu}}, \bibinfo {author} {\bibfnamefont {Z.}~\bibnamefont {Wang}}, \bibinfo {author} {\bibfnamefont {C.}~\bibnamefont {Zhao}}, \ and\ \bibinfo {author} {\bibfnamefont {Y.}~\bibnamefont {Cai}},\ }\href@noop {} {\bibfield  {journal} {\bibinfo  {journal} {Nanophotonics}\ }\textbf {\bibinfo {volume} {11}},\ \bibinfo {pages} {241} (\bibinfo {year} {2022})}\BibitemShut {NoStop}%
\bibitem [{\citenamefont {Ni}\ \emph {et~al.}(2021)\citenamefont {Ni}, \citenamefont {Huang}, \citenamefont {Zhou}, \citenamefont {Gu}, \citenamefont {Song}, \citenamefont {Kivshar},\ and\ \citenamefont {Qiu}}]{ni2021multidimensional}%
  \BibitemOpen
  \bibfield  {author} {\bibinfo {author} {\bibfnamefont {J.}~\bibnamefont {Ni}}, \bibinfo {author} {\bibfnamefont {C.}~\bibnamefont {Huang}}, \bibinfo {author} {\bibfnamefont {L.-M.}\ \bibnamefont {Zhou}}, \bibinfo {author} {\bibfnamefont {M.}~\bibnamefont {Gu}}, \bibinfo {author} {\bibfnamefont {Q.}~\bibnamefont {Song}}, \bibinfo {author} {\bibfnamefont {Y.}~\bibnamefont {Kivshar}}, \ and\ \bibinfo {author} {\bibfnamefont {C.-W.}\ \bibnamefont {Qiu}},\ }\href@noop {} {\bibfield  {journal} {\bibinfo  {journal} {Science}\ }\textbf {\bibinfo {volume} {374}},\ \bibinfo {pages} {eabj0039} (\bibinfo {year} {2021})}\BibitemShut {NoStop}%
\bibitem [{\citenamefont {Deng}\ \emph {et~al.}(2019)\citenamefont {Deng}, \citenamefont {Lin}, \citenamefont {Li},\ and\ \citenamefont {Zhao}}]{deng2019precision}%
  \BibitemOpen
  \bibfield  {author} {\bibinfo {author} {\bibfnamefont {D.}~\bibnamefont {Deng}}, \bibinfo {author} {\bibfnamefont {M.}~\bibnamefont {Lin}}, \bibinfo {author} {\bibfnamefont {Y.}~\bibnamefont {Li}}, \ and\ \bibinfo {author} {\bibfnamefont {H.}~\bibnamefont {Zhao}},\ }\href@noop {} {\bibfield  {journal} {\bibinfo  {journal} {Physical Review Applied}\ }\textbf {\bibinfo {volume} {12}},\ \bibinfo {pages} {014048} (\bibinfo {year} {2019})}\BibitemShut {NoStop}%
\bibitem [{\citenamefont {Lu}\ \emph {et~al.}(2025)\citenamefont {Lu}, \citenamefont {Yin}, \citenamefont {Zhao}, \citenamefont {Zhai},\ and\ \citenamefont {Wang}}]{lu2025fractional}%
  \BibitemOpen
  \bibfield  {author} {\bibinfo {author} {\bibfnamefont {Y.}~\bibnamefont {Lu}}, \bibinfo {author} {\bibfnamefont {Z.}~\bibnamefont {Yin}}, \bibinfo {author} {\bibfnamefont {W.}~\bibnamefont {Zhao}}, \bibinfo {author} {\bibfnamefont {A.}~\bibnamefont {Zhai}}, \ and\ \bibinfo {author} {\bibfnamefont {D.}~\bibnamefont {Wang}},\ }\href@noop {} {\bibfield  {journal} {\bibinfo  {journal} {Optics and Lasers in Engineering}\ }\textbf {\bibinfo {volume} {190}},\ \bibinfo {pages} {108956} (\bibinfo {year} {2025})}\BibitemShut {NoStop}%
\bibitem [{\citenamefont {Wang}\ \emph {et~al.}(2025{\natexlab{a}})\citenamefont {Wang}, \citenamefont {Yin}, \citenamefont {Sun}, \citenamefont {Xin}, \citenamefont {Zhang},\ and\ \citenamefont {Chang}}]{wang2025optical}%
  \BibitemOpen
  \bibfield  {author} {\bibinfo {author} {\bibfnamefont {L.}~\bibnamefont {Wang}}, \bibinfo {author} {\bibfnamefont {X.}~\bibnamefont {Yin}}, \bibinfo {author} {\bibfnamefont {Y.}~\bibnamefont {Sun}}, \bibinfo {author} {\bibfnamefont {X.}~\bibnamefont {Xin}}, \bibinfo {author} {\bibfnamefont {Q.}~\bibnamefont {Zhang}}, \ and\ \bibinfo {author} {\bibfnamefont {H.}~\bibnamefont {Chang}},\ }\href@noop {} {\bibfield  {journal} {\bibinfo  {journal} {Optics Express}\ }\textbf {\bibinfo {volume} {33}},\ \bibinfo {pages} {35694} (\bibinfo {year} {2025}{\natexlab{a}})}\BibitemShut {NoStop}%
\bibitem [{\citenamefont {Cui}\ \emph {et~al.}(2025)\citenamefont {Cui}, \citenamefont {Jin}, \citenamefont {Ding}, \citenamefont {Ren}, \citenamefont {Zhang}, \citenamefont {Tang}, \citenamefont {Yang}, \citenamefont {Zhang}, \citenamefont {Liu},\ and\ \citenamefont {Li}}]{cui2025recognition}%
  \BibitemOpen
  \bibfield  {author} {\bibinfo {author} {\bibfnamefont {Z.}~\bibnamefont {Cui}}, \bibinfo {author} {\bibfnamefont {Q.}~\bibnamefont {Jin}}, \bibinfo {author} {\bibfnamefont {J.}~\bibnamefont {Ding}}, \bibinfo {author} {\bibfnamefont {X.}~\bibnamefont {Ren}}, \bibinfo {author} {\bibfnamefont {H.}~\bibnamefont {Zhang}}, \bibinfo {author} {\bibfnamefont {Z.}~\bibnamefont {Tang}}, \bibinfo {author} {\bibfnamefont {X.}~\bibnamefont {Yang}}, \bibinfo {author} {\bibfnamefont {H.}~\bibnamefont {Zhang}}, \bibinfo {author} {\bibfnamefont {K.}~\bibnamefont {Liu}}, \ and\ \bibinfo {author} {\bibfnamefont {X.}~\bibnamefont {Li}},\ }\href@noop {} {\bibfield  {journal} {\bibinfo  {journal} {Applied Physics Letters}\ }\textbf {\bibinfo {volume} {127}} (\bibinfo {year} {2025})}\BibitemShut {NoStop}%
\bibitem [{\citenamefont {Berger}\ \emph {et~al.}(2018)\citenamefont {Berger}, \citenamefont {Kahlert}, \citenamefont {Schmidt},\ and\ \citenamefont {Assmann}}]{berger2018spectroscopy}%
  \BibitemOpen
  \bibfield  {author} {\bibinfo {author} {\bibfnamefont {B.}~\bibnamefont {Berger}}, \bibinfo {author} {\bibfnamefont {M.}~\bibnamefont {Kahlert}}, \bibinfo {author} {\bibfnamefont {D.}~\bibnamefont {Schmidt}}, \ and\ \bibinfo {author} {\bibfnamefont {M.}~\bibnamefont {Assmann}},\ }\href@noop {} {\bibfield  {journal} {\bibinfo  {journal} {Optics express}\ }\textbf {\bibinfo {volume} {26}},\ \bibinfo {pages} {32248} (\bibinfo {year} {2018})}\BibitemShut {NoStop}%
\bibitem [{\citenamefont {Zhang}\ \emph {et~al.}(2021)\citenamefont {Zhang}, \citenamefont {Yuan}, \citenamefont {Ding}, \citenamefont {Li}, \citenamefont {Ratni}, \citenamefont {Wu}, \citenamefont {Liu}, \citenamefont {Burokur},\ and\ \citenamefont {Tan}}]{zhang2021polarization}%
  \BibitemOpen
  \bibfield  {author} {\bibinfo {author} {\bibfnamefont {K.}~\bibnamefont {Zhang}}, \bibinfo {author} {\bibfnamefont {Y.}~\bibnamefont {Yuan}}, \bibinfo {author} {\bibfnamefont {X.}~\bibnamefont {Ding}}, \bibinfo {author} {\bibfnamefont {H.}~\bibnamefont {Li}}, \bibinfo {author} {\bibfnamefont {B.}~\bibnamefont {Ratni}}, \bibinfo {author} {\bibfnamefont {Q.}~\bibnamefont {Wu}}, \bibinfo {author} {\bibfnamefont {J.}~\bibnamefont {Liu}}, \bibinfo {author} {\bibfnamefont {S.~N.}\ \bibnamefont {Burokur}}, \ and\ \bibinfo {author} {\bibfnamefont {J.}~\bibnamefont {Tan}},\ }\href@noop {} {\bibfield  {journal} {\bibinfo  {journal} {Laser \& Photonics Reviews}\ }\textbf {\bibinfo {volume} {15}},\ \bibinfo {pages} {2000351} (\bibinfo {year} {2021})}\BibitemShut {NoStop}%
\bibitem [{\citenamefont {Zhu}\ \emph {et~al.}(2024)\citenamefont {Zhu}, \citenamefont {Wang},\ and\ \citenamefont {Zhao}}]{zhu2024security}%
  \BibitemOpen
  \bibfield  {author} {\bibinfo {author} {\bibfnamefont {J.}~\bibnamefont {Zhu}}, \bibinfo {author} {\bibfnamefont {L.}~\bibnamefont {Wang}}, \ and\ \bibinfo {author} {\bibfnamefont {S.}~\bibnamefont {Zhao}},\ }\href@noop {} {\bibfield  {journal} {\bibinfo  {journal} {ACS Photonics}\ }\textbf {\bibinfo {volume} {11}},\ \bibinfo {pages} {4626} (\bibinfo {year} {2024})}\BibitemShut {NoStop}%
\bibitem [{\citenamefont {Xu}\ \emph {et~al.}(2025)\citenamefont {Xu}, \citenamefont {Wu}, \citenamefont {Zhong}, \citenamefont {Wu}, \citenamefont {He}, \citenamefont {Liang}, \citenamefont {Feng}, \citenamefont {Wen},\ and\ \citenamefont {Chen}}]{xu2025multiplexed}%
  \BibitemOpen
  \bibfield  {author} {\bibinfo {author} {\bibfnamefont {Z.}~\bibnamefont {Xu}}, \bibinfo {author} {\bibfnamefont {L.}~\bibnamefont {Wu}}, \bibinfo {author} {\bibfnamefont {W.}~\bibnamefont {Zhong}}, \bibinfo {author} {\bibfnamefont {Z.}~\bibnamefont {Wu}}, \bibinfo {author} {\bibfnamefont {Q.}~\bibnamefont {He}}, \bibinfo {author} {\bibfnamefont {Z.}~\bibnamefont {Liang}}, \bibinfo {author} {\bibfnamefont {G.}~\bibnamefont {Feng}}, \bibinfo {author} {\bibfnamefont {Y.}~\bibnamefont {Wen}}, \ and\ \bibinfo {author} {\bibfnamefont {Y.}~\bibnamefont {Chen}},\ }\href@noop {} {\bibfield  {journal} {\bibinfo  {journal} {Laser \& Photonics Reviews}\ }\textbf {\bibinfo {volume} {19}},\ \bibinfo {pages} {2401954} (\bibinfo {year} {2025})}\BibitemShut {NoStop}%
\bibitem [{\citenamefont {Zhang}\ \emph {et~al.}(2025{\natexlab{a}})\citenamefont {Zhang}, \citenamefont {Qiu}, \citenamefont {Huang}, \citenamefont {Xu},\ and\ \citenamefont {Wang}}]{zhang2025optical}%
  \BibitemOpen
  \bibfield  {author} {\bibinfo {author} {\bibfnamefont {K.}~\bibnamefont {Zhang}}, \bibinfo {author} {\bibfnamefont {J.}~\bibnamefont {Qiu}}, \bibinfo {author} {\bibfnamefont {Z.}~\bibnamefont {Huang}}, \bibinfo {author} {\bibfnamefont {B.}~\bibnamefont {Xu}}, \ and\ \bibinfo {author} {\bibfnamefont {X.}~\bibnamefont {Wang}},\ }\href@noop {} {\bibfield  {journal} {\bibinfo  {journal} {Optics Letters}\ }\textbf {\bibinfo {volume} {50}},\ \bibinfo {pages} {7424} (\bibinfo {year} {2025}{\natexlab{a}})}\BibitemShut {NoStop}%
\bibitem [{\citenamefont {Shen}\ \emph {et~al.}(2024)\citenamefont {Shen}, \citenamefont {Zhang}, \citenamefont {Shi}, \citenamefont {Du}, \citenamefont {Yuan},\ and\ \citenamefont {Zayats}}]{shen2024optical}%
  \BibitemOpen
  \bibfield  {author} {\bibinfo {author} {\bibfnamefont {Y.}~\bibnamefont {Shen}}, \bibinfo {author} {\bibfnamefont {Q.}~\bibnamefont {Zhang}}, \bibinfo {author} {\bibfnamefont {P.}~\bibnamefont {Shi}}, \bibinfo {author} {\bibfnamefont {L.}~\bibnamefont {Du}}, \bibinfo {author} {\bibfnamefont {X.}~\bibnamefont {Yuan}}, \ and\ \bibinfo {author} {\bibfnamefont {A.~V.}\ \bibnamefont {Zayats}},\ }\href@noop {} {\bibfield  {journal} {\bibinfo  {journal} {Nature Photonics}\ }\textbf {\bibinfo {volume} {18}},\ \bibinfo {pages} {15} (\bibinfo {year} {2024})}\BibitemShut {NoStop}%
\bibitem [{\citenamefont {Ma}\ \emph {et~al.}(2025)\citenamefont {Ma}, \citenamefont {Yang}, \citenamefont {Liu}, \citenamefont {Chen}, \citenamefont {Li}, \citenamefont {Song}, \citenamefont {Qiu}, \citenamefont {Zou}, \citenamefont {Hu}, \citenamefont {Li} \emph {et~al.}}]{ma2025nanophotonic}%
  \BibitemOpen
  \bibfield  {author} {\bibinfo {author} {\bibfnamefont {J.}~\bibnamefont {Ma}}, \bibinfo {author} {\bibfnamefont {J.}~\bibnamefont {Yang}}, \bibinfo {author} {\bibfnamefont {S.}~\bibnamefont {Liu}}, \bibinfo {author} {\bibfnamefont {B.}~\bibnamefont {Chen}}, \bibinfo {author} {\bibfnamefont {X.}~\bibnamefont {Li}}, \bibinfo {author} {\bibfnamefont {C.}~\bibnamefont {Song}}, \bibinfo {author} {\bibfnamefont {G.}~\bibnamefont {Qiu}}, \bibinfo {author} {\bibfnamefont {K.}~\bibnamefont {Zou}}, \bibinfo {author} {\bibfnamefont {X.}~\bibnamefont {Hu}}, \bibinfo {author} {\bibfnamefont {F.}~\bibnamefont {Li}},  \emph {et~al.},\ }\href@noop {} {\bibfield  {journal} {\bibinfo  {journal} {Nature Physics}\ }\textbf {\bibinfo {volume} {21}},\ \bibinfo {pages} {1462} (\bibinfo {year} {2025})}\BibitemShut {NoStop}%
\bibitem [{\citenamefont {Koni}\ \emph {et~al.}(2025)\citenamefont {Koni}, \citenamefont {Nothlawala}, \citenamefont {Hakobyan}, \citenamefont {Nape}, \citenamefont {Brasselet},\ and\ \citenamefont {Forbes}}]{koni2025dual}%
  \BibitemOpen
  \bibfield  {author} {\bibinfo {author} {\bibfnamefont {M.}~\bibnamefont {Koni}}, \bibinfo {author} {\bibfnamefont {F.}~\bibnamefont {Nothlawala}}, \bibinfo {author} {\bibfnamefont {V.}~\bibnamefont {Hakobyan}}, \bibinfo {author} {\bibfnamefont {I.}~\bibnamefont {Nape}}, \bibinfo {author} {\bibfnamefont {E.}~\bibnamefont {Brasselet}}, \ and\ \bibinfo {author} {\bibfnamefont {A.}~\bibnamefont {Forbes}},\ }\href@noop {} {\bibfield  {journal} {\bibinfo  {journal} {arXiv preprint arXiv:2507.22815}\ } (\bibinfo {year} {2025})}\BibitemShut {NoStop}%
\bibitem [{\citenamefont {Ornelas}\ \emph {et~al.}(2024)\citenamefont {Ornelas}, \citenamefont {Nape}, \citenamefont {de~Mello~Koch},\ and\ \citenamefont {Forbes}}]{ornelas2024non}%
  \BibitemOpen
  \bibfield  {author} {\bibinfo {author} {\bibfnamefont {P.}~\bibnamefont {Ornelas}}, \bibinfo {author} {\bibfnamefont {I.}~\bibnamefont {Nape}}, \bibinfo {author} {\bibfnamefont {R.}~\bibnamefont {de~Mello~Koch}}, \ and\ \bibinfo {author} {\bibfnamefont {A.}~\bibnamefont {Forbes}},\ }\href@noop {} {\bibfield  {journal} {\bibinfo  {journal} {Nature Photonics}\ }\textbf {\bibinfo {volume} {18}},\ \bibinfo {pages} {258} (\bibinfo {year} {2024})}\BibitemShut {NoStop}%
\bibitem [{\citenamefont {Tsesses}\ \emph {et~al.}(2018)\citenamefont {Tsesses}, \citenamefont {Ostrovsky}, \citenamefont {Cohen}, \citenamefont {Gjonaj}, \citenamefont {Lindner},\ and\ \citenamefont {Bartal}}]{tsesses2018evanescent}%
  \BibitemOpen
  \bibfield  {author} {\bibinfo {author} {\bibfnamefont {S.}~\bibnamefont {Tsesses}}, \bibinfo {author} {\bibfnamefont {E.}~\bibnamefont {Ostrovsky}}, \bibinfo {author} {\bibfnamefont {K.}~\bibnamefont {Cohen}}, \bibinfo {author} {\bibfnamefont {B.}~\bibnamefont {Gjonaj}}, \bibinfo {author} {\bibfnamefont {N.~H.}\ \bibnamefont {Lindner}}, \ and\ \bibinfo {author} {\bibfnamefont {G.}~\bibnamefont {Bartal}},\ }\href@noop {} {\bibfield  {journal} {\bibinfo  {journal} {Proceedings of the Royal Society of London. Series A: Mathematical, Physical and Engineering Sciences}\ }\textbf {\bibinfo {volume} {361}},\ \bibinfo {pages} {993} (\bibinfo {year} {2018})}\BibitemShut {NoStop}%
\bibitem [{\citenamefont {Guti{\'e}rrez-Cuevas}\ and\ \citenamefont {Pisanty}(2021)}]{gutierrez2021optical}%
  \BibitemOpen
  \bibfield  {author} {\bibinfo {author} {\bibfnamefont {R.}~\bibnamefont {Guti{\'e}rrez-Cuevas}}\ and\ \bibinfo {author} {\bibfnamefont {E.}~\bibnamefont {Pisanty}},\ }\href@noop {} {\bibfield  {journal} {\bibinfo  {journal} {Journal of Optics}\ }\textbf {\bibinfo {volume} {23}},\ \bibinfo {pages} {024004} (\bibinfo {year} {2021})}\BibitemShut {NoStop}%
\bibitem [{\citenamefont {Du}\ \emph {et~al.}(2019)\citenamefont {Du}, \citenamefont {Yang}, \citenamefont {Zayats},\ and\ \citenamefont {Yuan}}]{du2019deep}%
  \BibitemOpen
  \bibfield  {author} {\bibinfo {author} {\bibfnamefont {L.}~\bibnamefont {Du}}, \bibinfo {author} {\bibfnamefont {A.}~\bibnamefont {Yang}}, \bibinfo {author} {\bibfnamefont {A.~V.}\ \bibnamefont {Zayats}}, \ and\ \bibinfo {author} {\bibfnamefont {X.}~\bibnamefont {Yuan}},\ }\href@noop {} {\bibfield  {journal} {\bibinfo  {journal} {Nature Physics}\ }\textbf {\bibinfo {volume} {15}},\ \bibinfo {pages} {650} (\bibinfo {year} {2019})}\BibitemShut {NoStop}%
\bibitem [{\citenamefont {Wang}\ \emph {et~al.}(2025{\natexlab{b}})\citenamefont {Wang}, \citenamefont {Che}, \citenamefont {Cheng}, \citenamefont {Tong}, \citenamefont {Shi}, \citenamefont {Shen}, \citenamefont {Bliokh},\ and\ \citenamefont {Zi}}]{wang2025topological}%
  \BibitemOpen
  \bibfield  {author} {\bibinfo {author} {\bibfnamefont {B.}~\bibnamefont {Wang}}, \bibinfo {author} {\bibfnamefont {Z.}~\bibnamefont {Che}}, \bibinfo {author} {\bibfnamefont {C.}~\bibnamefont {Cheng}}, \bibinfo {author} {\bibfnamefont {C.}~\bibnamefont {Tong}}, \bibinfo {author} {\bibfnamefont {L.}~\bibnamefont {Shi}}, \bibinfo {author} {\bibfnamefont {Y.}~\bibnamefont {Shen}}, \bibinfo {author} {\bibfnamefont {K.~Y.}\ \bibnamefont {Bliokh}}, \ and\ \bibinfo {author} {\bibfnamefont {J.}~\bibnamefont {Zi}},\ }\href@noop {} {\bibfield  {journal} {\bibinfo  {journal} {Nature}\ ,\ \bibinfo {pages} {1}} (\bibinfo {year} {2025}{\natexlab{b}})}\BibitemShut {NoStop}%
\bibitem [{\citenamefont {de~Mello~Koch}\ \emph {et~al.}(2025)\citenamefont {de~Mello~Koch}, \citenamefont {Lu}, \citenamefont {Ornelas}, \citenamefont {Nape},\ and\ \citenamefont {Forbes}}]{de2025quantum}%
  \BibitemOpen
  \bibfield  {author} {\bibinfo {author} {\bibfnamefont {R.}~\bibnamefont {de~Mello~Koch}}, \bibinfo {author} {\bibfnamefont {B.-Q.}\ \bibnamefont {Lu}}, \bibinfo {author} {\bibfnamefont {P.}~\bibnamefont {Ornelas}}, \bibinfo {author} {\bibfnamefont {I.}~\bibnamefont {Nape}}, \ and\ \bibinfo {author} {\bibfnamefont {A.}~\bibnamefont {Forbes}},\ }\href@noop {} {\bibfield  {journal} {\bibinfo  {journal} {APL Quantum}\ }\textbf {\bibinfo {volume} {2}} (\bibinfo {year} {2025})}\BibitemShut {NoStop}%
\bibitem [{\citenamefont {Ornelas}\ \emph {et~al.}(2025)\citenamefont {Ornelas}, \citenamefont {Nape}, \citenamefont {de~Mello~Koch},\ and\ \citenamefont {Forbes}}]{ornelas2025topological}%
  \BibitemOpen
  \bibfield  {author} {\bibinfo {author} {\bibfnamefont {P.}~\bibnamefont {Ornelas}}, \bibinfo {author} {\bibfnamefont {I.}~\bibnamefont {Nape}}, \bibinfo {author} {\bibfnamefont {R.}~\bibnamefont {de~Mello~Koch}}, \ and\ \bibinfo {author} {\bibfnamefont {A.}~\bibnamefont {Forbes}},\ }\href@noop {} {\bibfield  {journal} {\bibinfo  {journal} {Nature Communications}\ }\textbf {\bibinfo {volume} {16}},\ \bibinfo {pages} {2934} (\bibinfo {year} {2025})}\BibitemShut {NoStop}%
\bibitem [{\citenamefont {Peters}\ \emph {et~al.}(2025)\citenamefont {Peters}, \citenamefont {Everts}, \citenamefont {Kleine}, \citenamefont {Ornelas},\ and\ \citenamefont {Forbes}}]{peters2025seeing}%
  \BibitemOpen
  \bibfield  {author} {\bibinfo {author} {\bibfnamefont {C.}~\bibnamefont {Peters}}, \bibinfo {author} {\bibfnamefont {K.}~\bibnamefont {Everts}}, \bibinfo {author} {\bibfnamefont {T.}~\bibnamefont {Kleine}}, \bibinfo {author} {\bibfnamefont {P.}~\bibnamefont {Ornelas}}, \ and\ \bibinfo {author} {\bibfnamefont {A.}~\bibnamefont {Forbes}},\ }\href@noop {} {\bibfield  {journal} {\bibinfo  {journal} {arXiv preprint arXiv:2508.12305}\ } (\bibinfo {year} {2025})}\BibitemShut {NoStop}%
\bibitem [{\citenamefont {Guo}\ \emph {et~al.}(2025)\citenamefont {Guo}, \citenamefont {Peters}, \citenamefont {Mata-Cervera}, \citenamefont {Vetlugin}, \citenamefont {Guo}, \citenamefont {Zhang}, \citenamefont {Forbes},\ and\ \citenamefont {Shen}}]{guo2025topological}%
  \BibitemOpen
  \bibfield  {author} {\bibinfo {author} {\bibfnamefont {Z.}~\bibnamefont {Guo}}, \bibinfo {author} {\bibfnamefont {C.}~\bibnamefont {Peters}}, \bibinfo {author} {\bibfnamefont {N.}~\bibnamefont {Mata-Cervera}}, \bibinfo {author} {\bibfnamefont {A.}~\bibnamefont {Vetlugin}}, \bibinfo {author} {\bibfnamefont {R.}~\bibnamefont {Guo}}, \bibinfo {author} {\bibfnamefont {P.}~\bibnamefont {Zhang}}, \bibinfo {author} {\bibfnamefont {A.}~\bibnamefont {Forbes}}, \ and\ \bibinfo {author} {\bibfnamefont {Y.}~\bibnamefont {Shen}},\ }\href@noop {} {\bibfield  {journal} {\bibinfo  {journal} {arXiv preprint arXiv:2509.05727}\ } (\bibinfo {year} {2025})}\BibitemShut {NoStop}%
\bibitem [{\citenamefont {Zhang}\ \emph {et~al.}(2025{\natexlab{b}})\citenamefont {Zhang}, \citenamefont {Xie}, \citenamefont {Zhuang}, \citenamefont {Wu}, \citenamefont {Liu}, \citenamefont {Wu}, \citenamefont {Mihalache}, \citenamefont {Shen},\ and\ \citenamefont {Deng}}]{Zhang2025Topological}%
  \BibitemOpen
  \bibfield  {author} {\bibinfo {author} {\bibfnamefont {Z.}~\bibnamefont {Zhang}}, \bibinfo {author} {\bibfnamefont {X.}~\bibnamefont {Xie}}, \bibinfo {author} {\bibfnamefont {C.}~\bibnamefont {Zhuang}}, \bibinfo {author} {\bibfnamefont {B.}~\bibnamefont {Wu}}, \bibinfo {author} {\bibfnamefont {Z.}~\bibnamefont {Liu}}, \bibinfo {author} {\bibfnamefont {B.}~\bibnamefont {Wu}}, \bibinfo {author} {\bibfnamefont {D.}~\bibnamefont {Mihalache}}, \bibinfo {author} {\bibfnamefont {Y.}~\bibnamefont {Shen}}, \ and\ \bibinfo {author} {\bibfnamefont {D.}~\bibnamefont {Deng}},\ }\href@noop {} {\bibfield  {journal} {\bibinfo  {journal} {Photonics Research}\ }\textbf {\bibinfo {volume} {13}},\ \bibinfo {pages} {B1} (\bibinfo {year} {2025}{\natexlab{b}})}\BibitemShut {NoStop}%
\bibitem [{\citenamefont {Bezuidenhout}\ \emph {et~al.}(2025)\citenamefont {Bezuidenhout}, \citenamefont {Peters}, \citenamefont {Kumar}, \citenamefont {Forbes},\ and\ \citenamefont {Nape}}]{bezuidenhout2025deep}%
  \BibitemOpen
  \bibfield  {author} {\bibinfo {author} {\bibfnamefont {H.}~\bibnamefont {Bezuidenhout}}, \bibinfo {author} {\bibfnamefont {C.}~\bibnamefont {Peters}}, \bibinfo {author} {\bibfnamefont {R.}~\bibnamefont {Kumar}}, \bibinfo {author} {\bibfnamefont {A.}~\bibnamefont {Forbes}}, \ and\ \bibinfo {author} {\bibfnamefont {I.}~\bibnamefont {Nape}},\ }\href@noop {} {\bibfield  {journal} {\bibinfo  {journal} {arXiv preprint arXiv:2512.11115}\ } (\bibinfo {year} {2025})}\BibitemShut {NoStop}%
\bibitem [{\citenamefont {Wang}\ \emph {et~al.}(2025{\natexlab{c}})\citenamefont {Wang}, \citenamefont {Ma}, \citenamefont {Zhang}, \citenamefont {Zhao}, \citenamefont {Cai}, \citenamefont {Qiu}, \citenamefont {Dong},\ and\ \citenamefont {He}}]{wang2025perturbation}%
  \BibitemOpen
  \bibfield  {author} {\bibinfo {author} {\bibfnamefont {A.~A.}\ \bibnamefont {Wang}}, \bibinfo {author} {\bibfnamefont {Y.}~\bibnamefont {Ma}}, \bibinfo {author} {\bibfnamefont {Y.}~\bibnamefont {Zhang}}, \bibinfo {author} {\bibfnamefont {Z.}~\bibnamefont {Zhao}}, \bibinfo {author} {\bibfnamefont {Y.}~\bibnamefont {Cai}}, \bibinfo {author} {\bibfnamefont {X.}~\bibnamefont {Qiu}}, \bibinfo {author} {\bibfnamefont {B.}~\bibnamefont {Dong}}, \ and\ \bibinfo {author} {\bibfnamefont {C.}~\bibnamefont {He}},\ }\href@noop {} {\bibfield  {journal} {\bibinfo  {journal} {Nature Photonics}\ ,\ \bibinfo {pages} {1}} (\bibinfo {year} {2025}{\natexlab{c}})}\BibitemShut {NoStop}%
\bibitem [{\citenamefont {Dai}\ \emph {et~al.}(2020)\citenamefont {Dai}, \citenamefont {Zhou}, \citenamefont {Ghosh}, \citenamefont {Mong}, \citenamefont {Kubo}, \citenamefont {Huang},\ and\ \citenamefont {Petek}}]{dai2020plasmonic}%
  \BibitemOpen
  \bibfield  {author} {\bibinfo {author} {\bibfnamefont {Y.}~\bibnamefont {Dai}}, \bibinfo {author} {\bibfnamefont {Z.}~\bibnamefont {Zhou}}, \bibinfo {author} {\bibfnamefont {A.}~\bibnamefont {Ghosh}}, \bibinfo {author} {\bibfnamefont {R.~S.}\ \bibnamefont {Mong}}, \bibinfo {author} {\bibfnamefont {A.}~\bibnamefont {Kubo}}, \bibinfo {author} {\bibfnamefont {C.-B.}\ \bibnamefont {Huang}}, \ and\ \bibinfo {author} {\bibfnamefont {H.}~\bibnamefont {Petek}},\ }\href@noop {} {\bibfield  {journal} {\bibinfo  {journal} {Nature}\ }\textbf {\bibinfo {volume} {588}},\ \bibinfo {pages} {616} (\bibinfo {year} {2020})}\BibitemShut {NoStop}%
\end{thebibliography}

%

\end{document}